\documentclass[aps,twocolumn,pre,floatfix,showpacs,tightenlines]{revtex4}
\usepackage{epsfig,graphicx,times}
\usepackage{amstext}
\usepackage{amsmath}
\usepackage{amssymb}
\usepackage{graphicx}
\usepackage{latexsym}
\usepackage{bm}

\def\dbarit {{\mathchar'26\mkern-11mud}}

\begin{document}

\title{Quantum Thermodynamic Cycles and Quantum Heat Engines (II)}

\author{H. T. Quan}
\affiliation{Theoretical Division, MS B213, Los Alamos National Laboratory, Los Alamos,
NM, 87545, U.S.A.}

\begin{abstract}
We study the quantum mechanical generalization of force or pressure,
and then we extend the classical thermodynamic isobaric process to
quantum mechanical systems. Based on these efforts, we are able to
study the quantum version of thermodynamic cycles that consist of
quantum isobaric process, such as quantum Brayton cycle and quantum
Diesel cycle. We also consider the implementation of quantum Brayton
cycle and quantum Diesel cycle with some model systems, such as
single particle in 1D box and single-mode radiation field in a
cavity. These studies lay the microscopic (quantum mechanical)
foundation for Szilard-Zurek single molecule engine.
\end{abstract}

\maketitle

\pagenumbering{arabic}

\section{INTRODUCTION}

Quantum thermodynamics is the study of heat and work dynamics in
quantum mechanical systems \cite {quantum thermodynamics}. In the
extreme limit of small systems with only a few degrees of freedom,
both the finite-size effect and quantum effects influence the
thermodynamic properties of the system dramatically
\cite{phystoday05,scully03,quan06}. The traditional thermodynamic
theory based on classical systems of macroscopic size does not apply
any more, and the quantum mechanical generalization of
thermodynamics becomes necessary. The interplay between
thermodynamics and quantum physics has been an interesting research
topic since 1950s \cite{earlydays}. In recent years, with the
developments of nanotechnology and quantum information processing,
the study of the interface between quantum physics and
thermodynamics begins to attract more and more attention
\cite{moreattention}. Studies of quantum thermodynamics not only
promise important potential applications in nanotechnology and
quantum information processing, but also bring new insights to some
fundamental problems of thermodynamics, such as Maxwell's demon and
the universality of the second law \cite{maxwell}. Among all the
studies about quantum thermodynamics, a central concern is to make
quantum mechanical extension of classical thermodynamic processes
and cycles \cite{quan07}.

It is well know that in classical thermodynamics there are four
basic thermodynamic processes: adiabatic process, isothermal
processes, isochoric process, and isobaric process
\cite{fourprocesses}. These four processes correspond to constant
entropy, constant temperature, constant volume, and constant
pressure, respectively. From these four basic thermodynamic
processes, we can construct all kinds of thermodynamic cycles, such
as Carnot cycle, Otto cycle, Brayton cycle, et al \cite{atoz}. Among
all the four kinds of basic thermodynamic processes, adiabatic
process has been extended to quantum domain, and has been
extensively studied ever since the born of quantum mechanics.
Nevertheless no attention was paid to the quantum mechanical
generalization of the remaining three basic thermodynamic processes
until most recently. In a recent paper \cite{quan07}, along with our
collaborators, we systematically study the quantum mechanical
generalization of the isothermal and the isochoric process. Base on
these studies, the properties of quantum Carnot cycle and quantum
Otto cycle are clarified. Meanwhile in recent years, numerous
studies on other quantum thermodynamic cycles are also reported
\cite{numerousstudy}. However, as to our best knowledge, the quantum
mechanical generalization of isobaric process (constant pressure)
has not been studied systematically so far. Possibly the lack of the
consideration of quantum isobaric process is due to the fact that
``pressure" (force) \cite{pressure} is not a well defined observable
in a quantum mechanical system. Because of 
the short of a well defined ``pressure" (force) and thus
the quantum isobaric process, the properties of quantum
thermodynamic cycles that consist  of quantum isobaric process, such
as quantum Brayton cycle and quantum Diesel cycle
\cite{fourprocesses,atoz} cannot be clarified. We notice that some
discussions about quantum Brayton cycle have been reported
\cite{brayton}. Nevertheless, their definitions of quantum isobaric
process and quantum Brayton cycle are ambiguous, and even in
contradiction sometimes. As a result, in their studies they cannot bridge the
quantum and classical thermodynamic cycles.

In this paper, along with our previous effort \cite {quan07}, we
will focus on the study of the quantum isobaric process \cite{atoz}
and its related quantum thermodynamic cycles. We begin with the
definition of ``pressure" for an arbitrary quantum system, and then
generalize the isobaric process to quantum mechanical systems. Based
on  this and our previous \cite{quan07} generalizations of
thermodynamic processes, we are able to study an arbitrary quantum
thermodynamic cycle constructed by any of these four quantum
thermodynamic processes. As an example, we will discuss the quantum
Brayton cycle and the quantum Diesel cycle and compare their
properties with their classical counterpart. Comparisons between
these quantum thermodynamic cycles and their classical counterparts
enable us to extend our understanding about the thermodynamics at
the interface of classical and quantum physics. This paper is
organized as follows: In Sec. II, we define microscopically
``pressure" for an arbitrary quantum mechanical system and study the
quantum mechanical generalization of isobaric process; in Sec. III,
we study quantum Brayton cycle and study how the efficiency of
Brayton cycle bridges quantum and classical thermodynamic cycles; in
Sec. IV we study quantum Diesel cycle in comparison with their
classical counter part; Sec. V is the remarks and conclusion.

\section{QUANTUM ISOBARIC PROCESS}

\subsection{Pressure in quantum-mechanical system}
In order to study the quantum isobaric process, we must first study
pressure in an arbitrary quantum mechanical system. Let us recall
that in some previous work \cite {quan07, workandheat}, heat and
work have been extended to quantum mechanical systems and expressed
as functions of the eigenenergies $E_{n}$ and probability
distributions $P_{n}$. The first law of thermodynamics has also been
generalized to quantum mechanical systems:
\begin{eqnarray}
\begin{split}
\dbarit Q &=\sum_n E_n dP_n, \label{1}\\
\dbarit W &=\sum_n P_n dE_n, \label{2}\\
dU&=\dbarit Q+\dbarit W=\sum_n(E_n dP_n+P_n dE_n),
\end{split}
\end{eqnarray}
where $E_n$ is the $n$th eigenenergy of the quantum mechanical
system with the Hamiltonian $H=\sum_{n}E_{n}\left\vert
n\right\rangle \left\langle n\right\vert$ under consideration; $P_n$
is the occupation probability in the $n$th eigenstate; $\left\vert n
\right\rangle$ is the $n$th eigenstate of the Hamiltonian. The
density matrix of the system can be written as $\rho = \sum_{n}
P_{n}\left\vert n \right\rangle \left\langle n \right\vert$.
$\dbarit Q $ and $\dbarit W$ depict the heat exchange and work done
respectively during a thermodynamic process. From classical
thermodynamics we know that the first law can be expressed as
$dU=\dbarit Q+\dbarit W=TdS+\sum_n Y_n dy_n$. Here, $T$ and $S$
refer to temperature and the thermodynamic entropy; $Y_n$ is the
generalized force, and $y_n$ is generalized coordinate corresponding
to $Y_{n}$ ($dy_n$ is the generalized displacement) \cite
{generalizedforce}. Inversely, the generalized force conjugate to the generalized coordinate $y_{n}$ can be expressed as \cite{onsager}
\begin{equation}
Y_n=-\frac {\dbarit W} {dy_n}.
\end{equation}
For example, when the generalized coordinate is chosen to be the
volume $V$, we have its corresponding generalized force -- pressure
$P= - \dbarit W / dV$. Motivated by the definition of the
generalized force for a classical system, we define analogously the
force (for 1D system, force is the same as pressure) for a quantum
mechanical system
\begin {equation}
F=-\frac{\dbarit W}{dL}=-\sum_n P_n \frac{dE_n}{dL}, \label{3}
\end{equation}
where $L$ is the generalized coordinate corresponding to the force
$F$. In obtaining Eq. (3), we have used the expression of work for a quantum system $\dbarit W =\sum_n P_n dE_n$. For a single particle in a 1D box (1DB) \cite{szilard} (see
Fig. 1), the generalized coordinate is the width of the potential, and the eigenenergies for such a system depend on the
generalized coordinate $E_n(L)=(\pi\hbar n)^2/(2mL^2)$. Here $\hbar$ is the Plank's constant; 
$n$ is the quantum number; $m$ is the mass of the particle. We
obtain the derivative of $E_n(L)$ over $L$ straightforwardly $\frac{dE_n}{dL}=-2 \frac{E_n(L)}{L}$. When the system is in thermal
equilibrium with a heat bath at a inverse temperature $\beta=\frac{1}{kT}$, the force
exerting on either wall of the potential can be calculated by
substituting $\frac{dE_n}{dL}$ and the Gibbs distribution
$P_{n}=\frac{1}{Z} e^{- \beta E_{n}}$ into Eq. (\ref{3}). Here $k$ is the Boltzmann's constant, and
$Z=\sum_{n} e^{-\beta E_{n}}$ is the partition function. Alternatively, the expression of force (3)
in a quantum mechanical system can be obtained in a
statistical-mechanical way \cite{pathria}.
\begin{equation}
\begin{split}
F&=-\left(\frac{\partial \mathbb{F} }{\partial L}\right)_{T} =kT \left(\frac{\partial  \ln{Z}  }{\partial L}\right)_{T}=kT \frac{1}{Z} \frac{\partial}{\partial L}\sum_{n} e^{-\beta E_{n}}\\
  &=-\sum_{n}\left(\frac{e^{-\beta E_{n}}}{Z}\right) \frac{\partial E_{n}}{\partial L}=-\sum_{n}P_{n}\frac{d E_{n}}{d L}, \label{4}
\end{split}
\end{equation}
where $\mathbb{F}=-kT \ln{Z}$ is the free energy of the quantum
system. It should be pointed out that Eq. (\ref{3}) is more general
than Eq. (4) because Eq. (\ref{3}) stands no matter the system is in
equilibrium or not. When $P_{n}$ in Eq. (\ref{3}) satisfies Gibbs
distribution $P_{n}=\frac{1}{Z} e^{- \beta E_{n}}$, or the system is
in thermal equilibrium, the expectation value of force $F$ of Eq. (3)
reproduce the usual force in classical thermodynamics.

Another model example is the single particle in a 1D harmonic
oscillator potential (1DH) (see Fig. 2). Its Hamiltonian is the same
as the Hamiltonian of a single-mode radiation field in a cavity
\cite{scully03}. For such 1DH, we will see later that the definition
of force (\ref{3}) for 1DH agrees with the radiation pressure (4) of
a single mode radiation field. We would like to mention that the
definition of force in (3) is a further step in quantum
thermodynamics after the definitions of heat and work (\ref {1}). We
will see all these definitions of work, heat, entropy and pressure for a
quantum mechanical system are  self-consistent, and consistent with
classical thermodynamics.
\begin{figure}[ht]
\begin{center}
\includegraphics[width=8cm, clip]{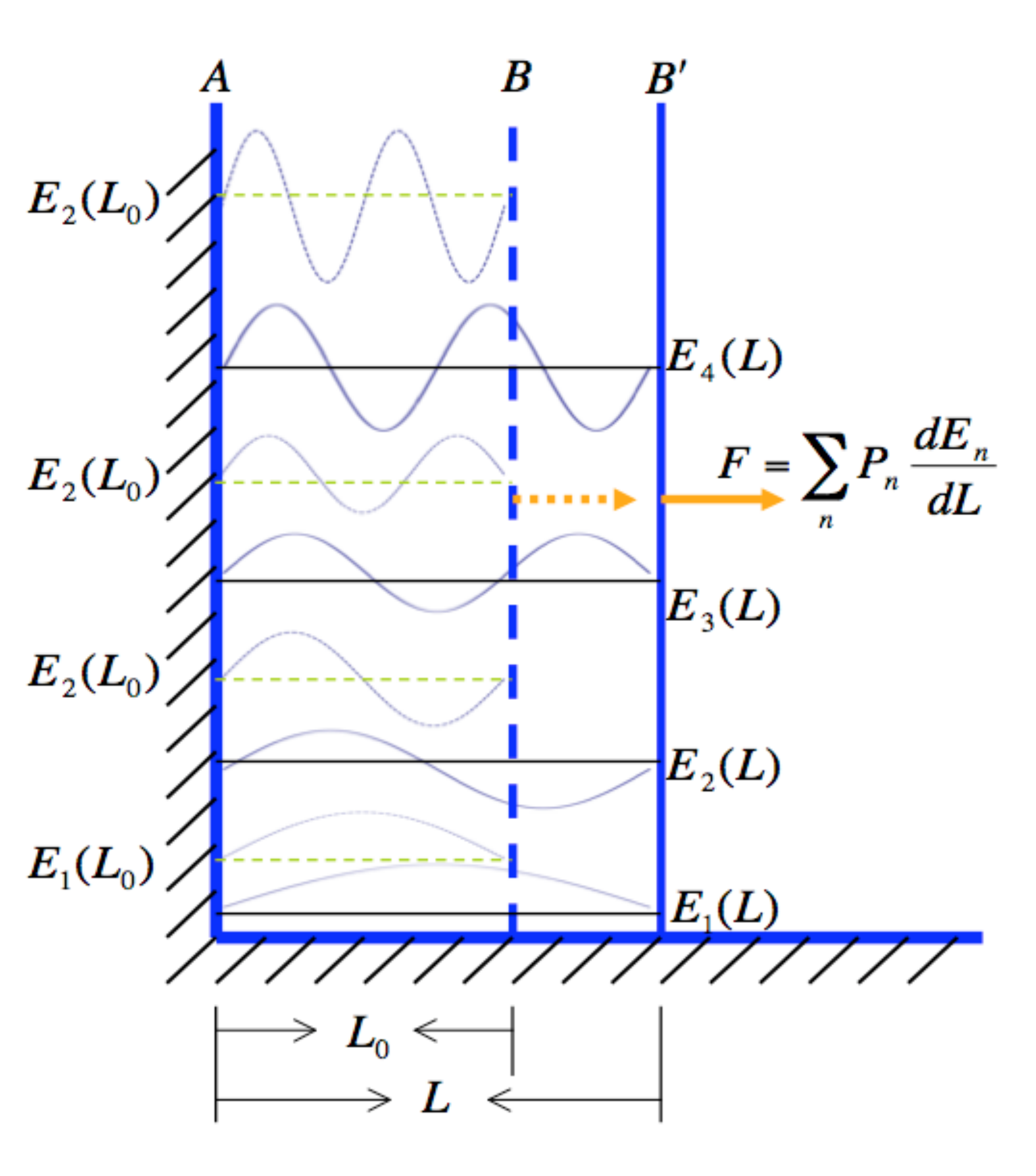}
\end{center}
\caption{Schematic diagram of  pressure in quantum mechanical system
(single particle in 1D box). One wall (A) of the square
well is fixed, while the other one (B) is movable. The force acting
on the wall B by the quantum system constrained in the potential well can be calculated from Eq.
(\ref {3})} \label{fig1}
\end{figure}

\subsection{Quantum isobaric process}

\begin{figure}[ht]
\begin{center}
\includegraphics[width=8cm, clip]{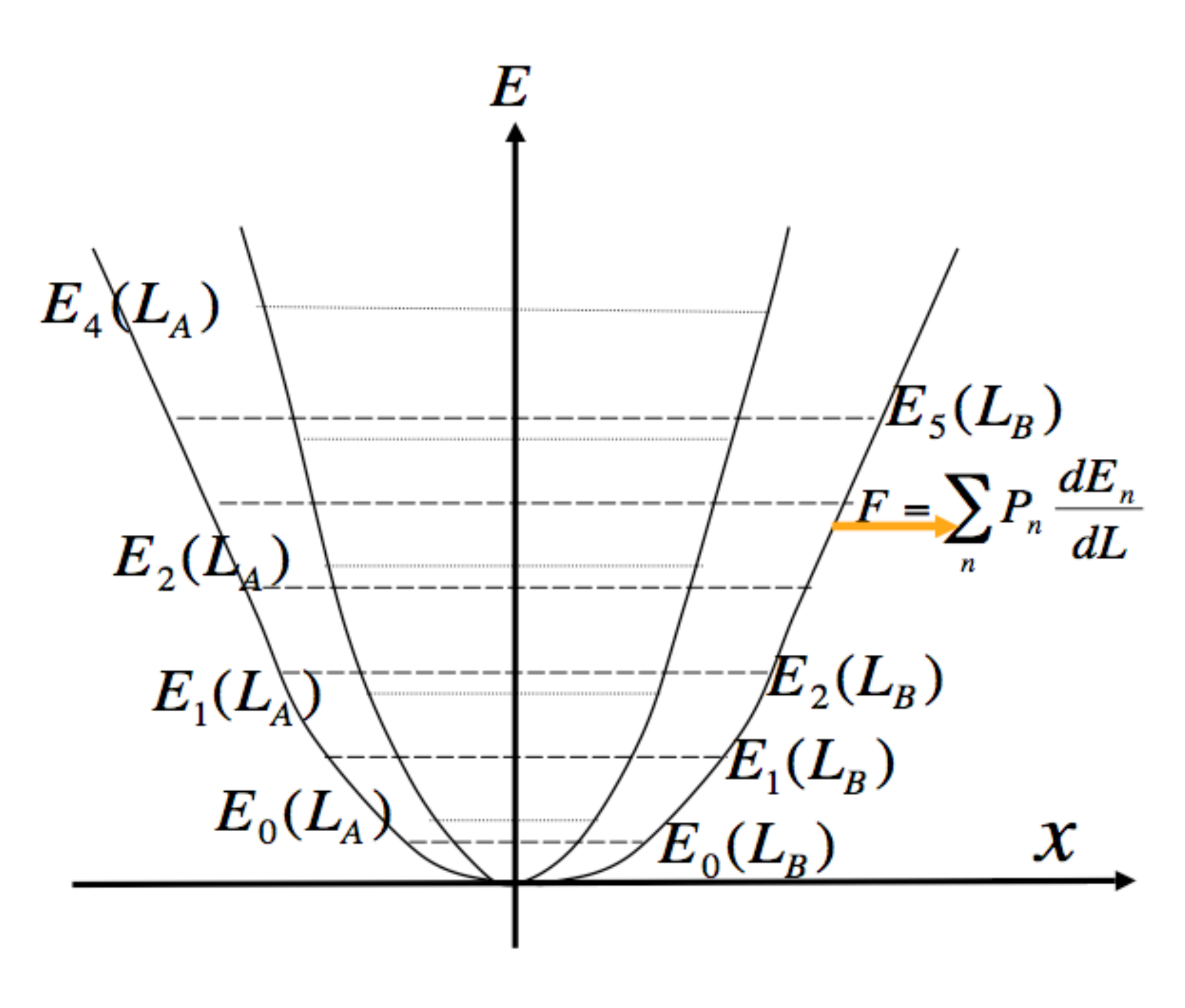}
\end{center}
\caption{Radiation pressure of a single-mode radiation field.
Here the width of the potential is inversely proportional to the mode frequency of the cavity
$L\propto \frac{1}{\omega}$. The $n$th eigenenergy of the
single mode radiation field with the potential width $L_{\alpha}$ is given by $E_{n}(L_{\alpha})=(n+\frac{1}{2})\frac{\hbar s\pi
c}{L_{\alpha}}$, $\alpha=A, B$. This pressure is equivalent to the pressure in a quantum mechanical system -- single particle
in 1D harmonic oscillator. Hence quantum isobaric process based on single-mode radiation field
is equivalent to that based on a 1D quantum harmonic oscillator.} \label{fig8}
\end{figure}
Having clarified force for a quantum mechanical system, in the
following we study how to extend classical isobaric process to a
quantum mechanical system. Classical isobaric process is a
quasi-static thermodynamic process, in which the pressure of the
system remains a constant \cite{fourprocesses,atoz}. The time scale
of relaxation of the system with the heat bath is much shorter than
the time scale of controlling the volume of the system \cite
{relaxation}. In a classical isobaric process, in order to achieve a
constant pressure, we must carefully control the temperature of the
system, i.e., carefully control the temperature of the heat bath,
when we change the volume of the classical system \cite {atoz}. For
example, for the classical idea gas with the equation of state
$PV=NkT$, the temperature of the system in the isobaric
process is required to be proportional to the volume$(T\propto V)$
of the gas, so that the pressure can remains a constant. For a quantum mechanical system, however, the change of
the temperature of the heat bath with the generalized coordinate may
not be so obvious as the classical ideal gas. Because we usually do not know the equation of state
of a quantum mechanical system. Let us consider the
quantum isobaric process based on 1DB (see Fig. 1). For such a
quantum mechanical system, the pressure on the wall can be obtained
from Eq. (3)
\begin{equation}
\begin{split}
F&=-\sum_n P_{n}(L) \frac {dE_n(L)}{dL} \label{4}\\
&=-\sum_n \frac {\exp[-\beta(L)E_n(L)]}{Z(L)}\frac {dE_n(L)}{dL}\\
&=-\sum_n \frac{\exp[-\beta(L)\frac{\pi^2\hbar^2n^2}{2mL^2}]}{\frac{1}{2}\sqrt{\frac{2mL^2}{\pi\hbar^2\beta(L)}}}\times \frac{(-2)}{L}\times \frac{\pi^2\hbar^2n^2}{2mL^2} \\
&=\frac{4}{L} \sqrt{\frac{\pi\hbar^2\beta(L)}{2mL^2}}  \left[-\frac{\partial}{\partial \beta(L)}\sum_n \exp[-\beta(L)\frac{\pi^2\hbar^2n^2}{2mL^2}] \right] \\
&=\frac{4}{L} \sqrt{\frac{\pi\hbar^2\beta(L)}{2mL^2}} \left[-\frac{\partial}{\partial \beta(L)} \frac{1}{2} \sqrt{\frac{2mL^2}{\pi\hbar^2\beta(L)}} \right] \\
&=\frac{1}{L\beta(L)}.
\end{split}
\end{equation}
Eq. (5) can be regarded as the equation of state $FL=kT$ for 1DB obtained from Eq. (3), and it means that if we want to keep the pressure $F$ as a
constant, we must control the temperature of the system to be
proportional to the width of the potential well $\beta(L) = 1/(FL)$ when the system inside the box pushes one of the walls
to perform work. This property of 1DB is the same as the classical ideal gas. We will see more analogues between them later. It should be mentioned that the temperature
function $\beta(L)$ of the ``volume" in a quantum isobaric process
is system-dependent. I.e., for different quantum systems, the
function of the temperature over the ``volume" in the quantum
isobaric process differs from one to another. In the following we
consider the quantum isobaric process based on a single mode
radiation field in a cavity, which was first proposed as the working
substance for a quantum heat engine in Ref. \cite{scully03}. We
assume that the cavity of length $L$ and cross-section $A$ can
support only a single mode of the field $\omega=\frac{s \pi c}{L}$,
where $s$ is an integer, and $c$ is the speed of light. The Hamiltonian reads
\begin{equation}
H=\sum_n(n+\frac{1}{2})\hbar\omega\left\vert n \right\rangle
\left\langle n \right \vert, \label {21}
\end{equation}
where $\left\vert n \right\rangle$ is the Fock state of the radiation field.
From Eq. (\ref{3}) we obtain the radiation force $F$ as a function of the temperature $\beta$ and the length of the cavity $L$ \cite{casimir}
\begin{equation}
\begin{split}
F&=-\sum_n\frac{e^{-\beta(L)E_n(L)}}{Z(L)}\frac{dE_n(L)}{dL}\\
&=-\frac{1}{1-e^{-\beta(L)\hbar\omega}}\sum_n e^{-\beta(L)n\hbar\omega}\left[(n+\frac{1}{2})\hbar\omega\right]\frac{1}{L},\label {22}\\
&=\left[\frac{\hbar\frac{s\pi c}{L}}{e^{\beta(L)\hbar\frac{s\pi
c}{L}}-1}+\frac{1}{2}\hbar\frac{s\pi c}{L} \right]\frac{1}{L}.
\end{split}
\end{equation}
From Eq. (7) it can be inferred that in order to achieve a constant
force, we must carefully control the temperature of the heat bath in
the following subtle way
\begin{equation} \beta (L)=\frac{L}{\hbar s \pi c}\ln{\frac{2 F
L^{2}+\hbar s \pi c}{2 F L^{2}-\hbar s \pi c}}.
\end{equation}
It can be seen that in a quantum isobaric process, the temperature
function (8) for the single-mode radiation field is much more
complicated than that $(\beta(L) \propto \frac{1}{L})$ of 1DB.

For the convenience of later analysis, we would also like to
calculate the entropy and the internal energy of the two systems in
a quantum isobaric process. First we consider the 1DB. The entropy
expression can be obtained from the above Eq. (\ref{1}) \cite
{quan07}.
\begin{equation}
\begin{split}
S(L)&=k_B\left[\frac{1}{2}+\ln\left(
\frac{1}{2}\sqrt{\frac{2mL^2}{\pi\hbar^2\beta(L)}} \right)\right].
\label {4.5}
\end{split}
\end{equation}
Through the comparison with the entropy of classical ideal gas
\cite{fourprocesses,atoz}, we find that the entropy of classical
idea gas reproduces the entropy (9) of 1DB if we choose the molecule
number of the classical ideal gas to be $N=1$. We plot the
entropy-temperature curve (\ref {4.5}) of a quantum isobaric process
in Fig. 3. The internal energy of the 1DB during the isobaric
process can also be obtained analytically (the temperature of the
heat bath is time-dependent).
\begin{equation}
U(L)=-\sum_n \frac{e^{- \beta (L) E_{n}(L)} }{Z(L)}  E_n(L)
\label{4}=\frac{1}{2\beta(L)}.
\end{equation}
This expression of internal energy verifies the equipartition
theorem \cite {fourprocesses}, and justifies the result in Ref.
\cite{quan07} again: the internal energy of the 1DB depends only on
the temperature. From Eqs. (9) and (10) we see that both the entropy
and the internal energy of 1DB have the same form as that of the
classical ideal gas \cite{fourprocesses,atoz} if we choose the molecule number of the classical ideal gas to be $N=1$. Moreover, from Eq.
(5) we know that 1DB has the same equation of state $FL=kT$ as that of the
classical ideal gas $PV=N k T$ except the difference of the particle number. Thus we conclude that 1DB is the
quantum mechanical counterpart of the classical ideal gas. We would
like to mention that in the study of the single-molecule engine by
Szilard and Zurek \cite{szilard}, they simply employ the equation of
state for the classical ideal gas $PV=N k_{B} T$ and choose the particle number to be $N=1$.
Nevertheless, this treatment may be questionable because the
equation of state $PV=N k_{B} T$ stands in the macroscopic and
classical case. When we come to the extreme limit of small system
with only a few degrees of freedom, we must use the quantum
mechanical treatment as we present here. Fortunately, all the
treatments of the single molecule engine \cite{szilard} by Szilard
and Zurek is in accordance with our quantum mechanical treatments.
Thus we say that our discussions lay the foundation for
Szilard-Zurek single molecule engine \cite{szilard}.

As to the single-mode radiation field, the entropy and the internal
energy can be calculated as that in Ref. \cite{scully03}
\begin{equation}
S(L)=\frac{\left\langle n\right\rangle \hbar \omega}{T}+k
\ln(\left\langle n\right\rangle+1)
\end{equation}
\begin{equation}
U(L)=\sum_n\frac{e^{-\beta n \hbar
\omega}}{Z(L)}(n+\frac{1}{2})\hbar \omega \label {23}=(\left\langle
n\right \rangle+\frac{1}{2})\hbar \omega
\end{equation}
where $\left\langle n\right \rangle=[\exp {(\hbar \omega
/kT)}-1]^{-1}$ is the mean photon number.

It is easy to see that the entropy (11) and the internal energy (12)
of a single mode radiation field have different forms from that of
1DB (9),(10), and thus from the classical ideal gas. The internal
energy (12) of single mode radiation field depends on both the
temperature $\beta$ and the width $L$ of the potential well, while
the internal energy of 1DB (10) depends on $\beta$ only. In
addition, the equation of state (7) of the single-mode radiation
field differs from from that (5) of 1DB, and thus from the classical
ideal gas. Based on these observations, we say that the single mode
radiation field has totally different thermodynamic properties from
that of classical ideal gas. It can be inferred that quantum heat
engine based on single mode radiation field can give us new results
beyond that of classical ideal gas. As we mentioned before, the
Hamiltonian of the single-mode radiation field is the same as that
of 1DH. Thus all the results about single mode radiation field are the
same as that for 1DH. Alternatively, we can say that 1DH is the counterpart of single-mode photon gas, in analogy to the
fact that 1DB is the counterpart of classical ideal gas. But it should be mentioned that single-mode
photon  gas are still quantum mechanical system, while classical ideal gas are classical system. In
the following we will alternatively use 1DH and single-mode photon
gas.

In addition to our previous studies \cite{quan07}, up to now we have
extended all four basic thermodynamic processes to quantum
mechanical domain. For a comparison of quantum thermodynamics
processes and their classical counterparts see Table I.

\begin{table*}[tbp]
\caption{Basic classical thermodynamic processes and their quantum
counterparts. Here the classical thermodynamic processes are based
on classical ideal gas, while the quantum thermodynamic processes
are based on the 1DB. We illustrate the equations of state for the
four basic thermodynamic processes, and we also indicate the
invariant or varying variables in these processes. Here, we use
``VRA" to indicate the invariance of a thermodynamic quantity and
``VAR" to indicate it varies or changes.}
\begin{center}
\begin{tabular}{c|c|c|c|c}
\hline\hline
& \parbox{3.5cm} { \textbf{Isothermal }($T\equiv T_0$)} & \parbox{3.6cm} {\textbf{Isochoric} \ \ \ \ \ \ \ \ \ \ \ \ \ \ \ \ \ ($V\equiv V_0$ or $L\equiv L_0$)} & \parbox{3.5cm} {\textbf{Isobaric} \ \ \ \ \ \ \ \ \ \ \ \ \ \ \ \ \ ($P\equiv P_0$ or $F\equiv F_0$)} & \parbox{3.5cm} {\textbf{Adiabatic} ($S\equiv S_0$)}\\
 \hline \hline
\textbf{Classical} &\parbox{3.5cm} {$P(V)V=const$; \ \ \ \ \ \ \ \ \ \ \ \ \ \ \ \ \  VRA: S, V, P; INV: T} &\parbox{3.8cm} {$\frac{P(T)}{T}=const$; \ \ \ \ \ \ \ \ \ \ \ \ \ \ \ \ \ VRA: S, T, P; INV: V}&\parbox{4.2cm} {$\frac{V(T)}{T}=const$; \ \ \ \ \ \ \ \ \ \ \ \ \ \ \ \ \ \ \ \ \ \ \ \ \ \ \ \ \ \ \ \ \ \  VRA: S, T, V; INV: P,}&\parbox{4.2cm} {$P(T)V^3(T)=const$; \ \ \ \ \ \ \ \ \ \ \ \ \ \ \ \ \ VRA: V, T P; INV: S}\\
\hline
\textbf{Quantum} &\parbox{3.5cm} {$F(L)L=const$; \ \ \ \ \ \ \ \ \ \ \ \ \ \ \ \ \  VRA:  $E_{n}$, $P_{n}$; INV: T} &\parbox{3.8cm} {$\frac{F(T)}{T}=const$; \ \ \ \ \ \ \ \ \ \ \ \ \ \ \ \ \ VRA:  T, $P_{n}$; INV: $E_{n}$}&\parbox{4.2cm} {$\frac{L(T)}{T}=const$;  \ \ \ \ \ \ \ \ \ \ \ \ \ \ \ \ \ \ \ \ \ \ \ \ \ \ \ \ \ \ \ \ \ \  VRA: T, $E_{n}$, $P_{n}$; }& \parbox{4.2cm} {$F(T)L^3(T)=const$; \ \ \ \ \ \ \ \ \ \ \ \ \ \ \ \ \ VRA: $E_{n}$, T; INV: , $P_{n}$}\\
 \hline\hline
\end{tabular}%
\end{center}
\end{table*}

\section{QUANTUM BRAYTON CYCLE}

\begin{figure}[ht]
\begin{center}
\includegraphics[width=8cm, clip]{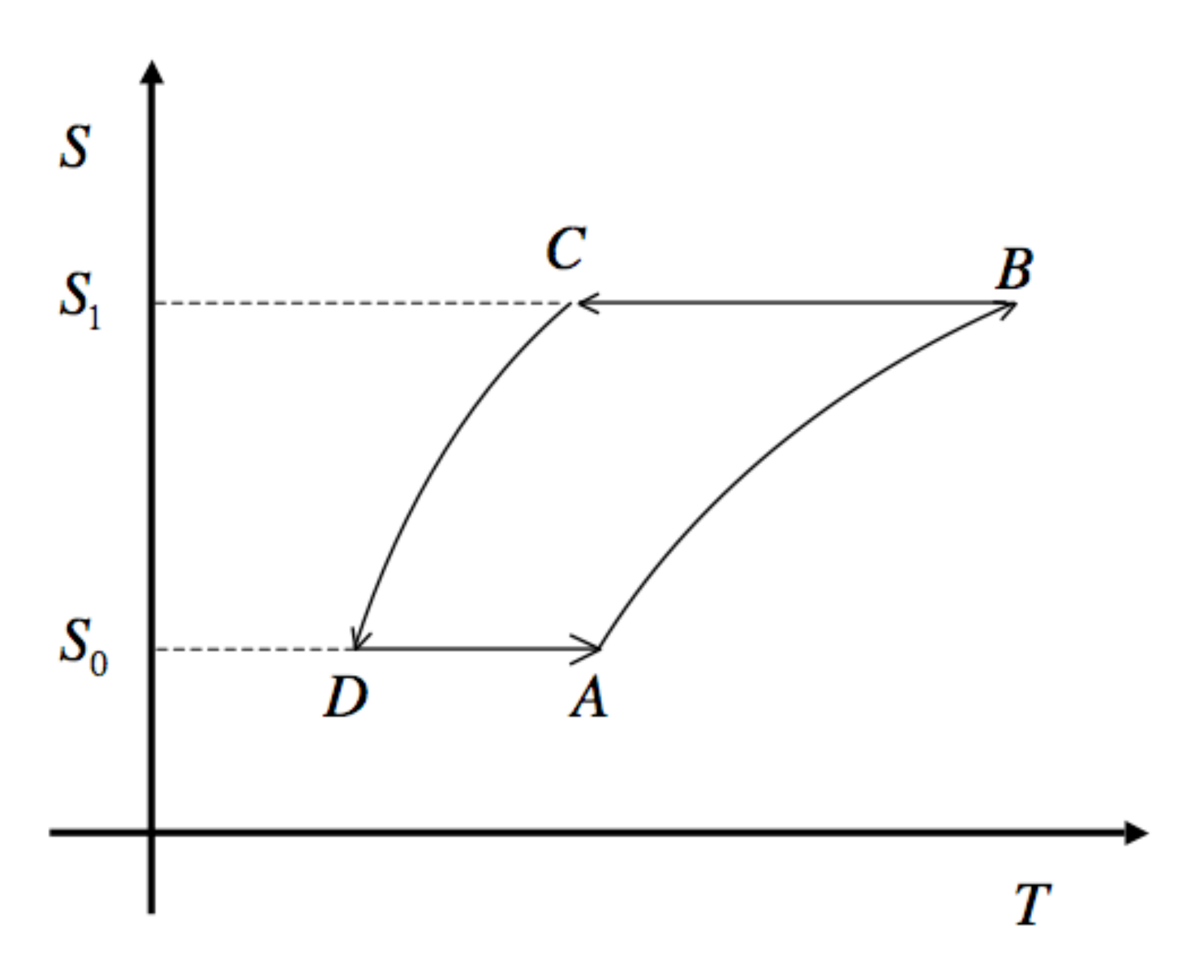}
\end{center}
\caption{Temperature-Entorpy $T-S$ diagram of a quantum Brayton
cycle (see Fig. 4) based on 1DB. In two adiabatic processes,
$B\longrightarrow C$ and $D \longrightarrow A$ the entropy remains a
constant. } \label{fig3}
\end{figure}

In the preceding section, we extend the classical isobaric process
to quantum mechanical systems based on the definition of pressure
(3). In this section and the next section, we study two kinds of
thermodynamic cycles consisting of the quantum isobaric process, and
compare them with their classical counterparts. We first consider
the quantum Brayton cycle based on 1DB. A quantum Brayton cycle is a
quantum mechanical analogue of the classical Brayton cycle
\cite{fourprocesses,atoz}, which consists of two quantum isobaric
processes and two quantum aidabatic processes. 
Similar to our discussion in Ref. \cite {quan07}, the counterpart of classical
adiabatic plus quasistatic process is quantum adiabatic process \cite{adiabatic}. In constructing
quantum Brayton cycle, we also requires that i) all the energy level spacing of the work substance
change by the same ratio in the quantum adiabatic process, and ii) this
ratio be equal to the ratio of the temperatures of the two heat baths just before and after
the quantum adiabatic process. It should be mentioned that in the isothermal process of a Carnot cycle,
the temperature of the heat bath is fixed. However, this is not the case in the isobaric process of a Bayton cycle.
Hence, we cannot simply say that the ratio of the change of the energy level spacings should be equal to the
ratio of the temperatures of the two heat baths \cite{quan07}. For the current example, it can be checked that
the change of the energy level spacings in quantum adiabatic process (see Fig. 3) $B \rightarrow C$ and 
$D \rightarrow A$ should be equal to the ratio of temperatures of the heat baths at $B$ and $C$, or at 
$D$ and $A$ (because $\frac{T_{B}}{T_{C}}=\frac{T_{A}}{T_{D}}$). 
Fortunately, the above condition i) can be satisfied by some quantum mechanical system,
such as 1DB, 1DH, two-level system, et al, and our study will focus on these systems whose energy level
spacings change in the same ratio in the quantum adiabatic process. Otherwise, the irreversibility will arise \cite{adiabatic}. 
We give a Temperature-Entropy $T-S$
diagram of the quantum Brayton cycle (See Fig. 3). Through a standard procedure, we obtain (see Appendix A) the
efficiency of the quantum Brayton cycle based on the 1DB
\begin{equation}
\eta_{\mathrm{Brayton}}=1-\left(\frac{F_0}{F_1}\right)^{\frac{2}{3}}, \label {20}
\end{equation}
where $F_{0}$ and $F_{1}$ are the pressures of the system during two quantum isobaric processes (See Fig. 4).
\begin{figure}[ht]
\begin{center}
\includegraphics[width=8cm, clip]{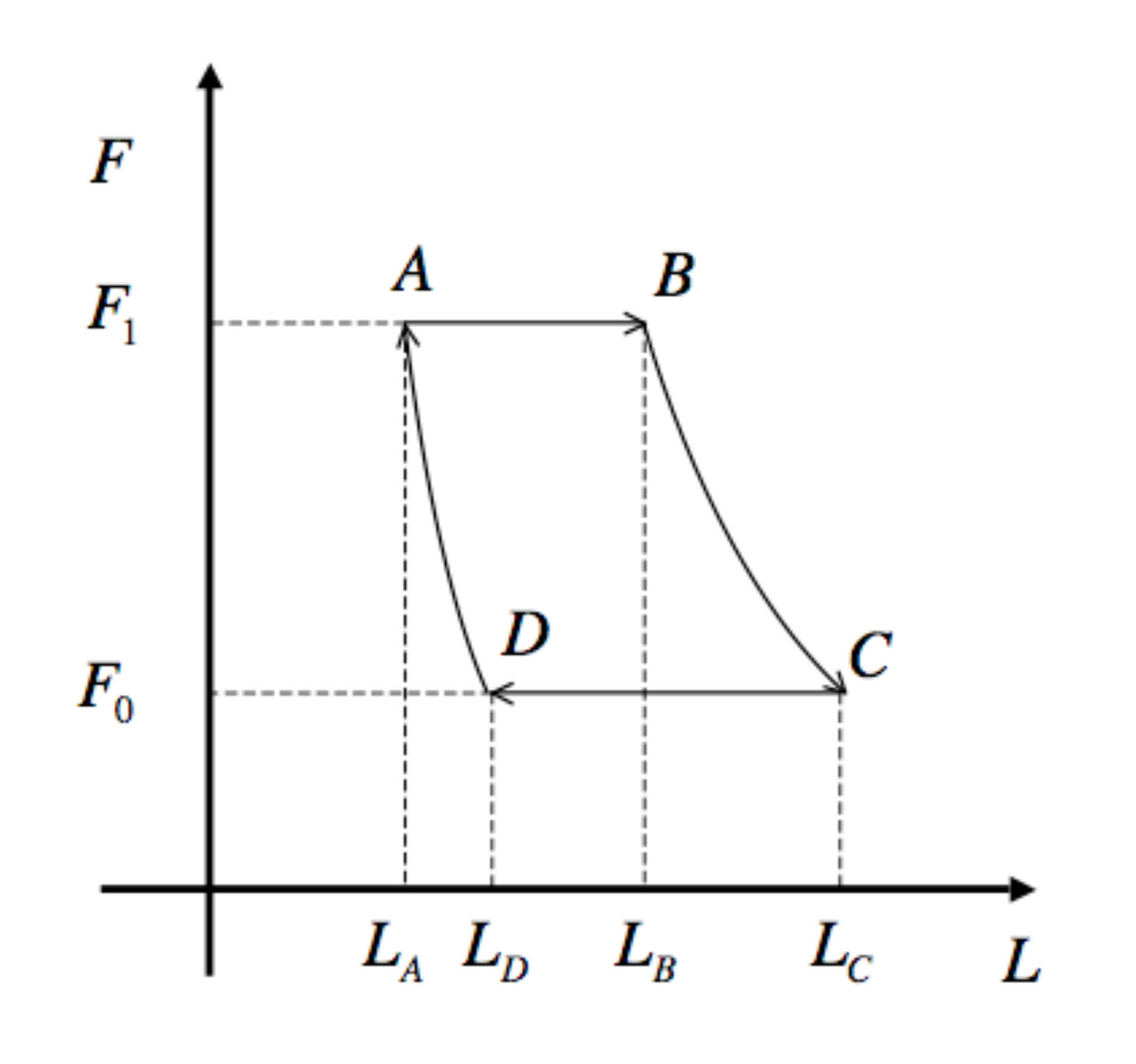}
\end{center}
\caption{Force-Displacement $F-L$ diagram of a quantum Brayton cycle
based on a single particle in 1DB. $A\longrightarrow B$ represents
an isobaric expansion process with a constant force $F_1$;
$B\longrightarrow C$ represents an adiabatic expansion process with
constant entropy $S_{1}$; $C\longrightarrow D$ represents an
isobaric compression process with constant pressure $F_0$;
$D\longrightarrow A$ is another adiabatic compression process with
constant entropy $S_{0}$.} \label{fig2}
\end{figure}

We would like to compare this efficiency of the quantum Brayton
cycle (13) with its classical counterpart. From Eq. (9) we know that
in the quantum adiabatic process ($S=const$), we have $TL^{2}=const$. As
a result the adiabatic exponent $\gamma=3$ is obtained through the
comparison with $TL^{\gamma-1}=const$ for the adiabatic process. Let us
recall that the efficiency of a classical Brayton cycle is
$\eta=1-\left(\frac{F_{0}}{F_{1}}\right)^{1-\frac{1}{\gamma}}$
\cite{atoz}, where $\gamma$ is the classical adiabatic exponent.
Thus our result (13) bridges the quantum Brayton cycle and classical
Brayton cycle. Hence this justifies that the definition of
pressure (3) for a quantum mechanical system is self-consistent.

Similarly we obtain we obtain the efficiency of a quantum Brayton cycle based on 1DH (see Appendix A)
\begin{equation}
\eta^{\prime}_{\mathrm{Brayton}}=1-\sqrt{\frac{F_0}{F_1}}. \label {31}
\end{equation}

From Eq. (11) we know that in a quantum adiabatic process
$TL=const$. Thus $\gamma=2$ for 1DH is obtained. It can be seen that
the efficiency of a quantum Brayton cycle obtained here (14) is the
same as that of a classical Brayton cycle. Through the discussion of
quantum Brayton cycles based on two model systems 1DH and 1DB, we
see that the definition of pressure (3) for quantum systems has
clear physical implication, and our study bridges the thermodynamic
cycles based on quantum and classical systems.

\section{QUANTUM DIESEL CYCLE}

Except for the thermodynamic cycles consisting of two pairs of
basic thermodynamic processes, such as Carnot cycle, Otto cycle \cite{quan07}, and Brayton cycle, there are some interesting
thermodynamic cycles consisting of more than two kinds of
thermodynamic processes, such as Diesel cycle. The Diesel cycle
consists of two adiabatic processes, one isobaric processes and one
isochcoric process \cite{atoz} (see Fig. 5). In order to construct such a 
quantum Diesel cycle, we require 1) the quantum adiabatic conditions are satisfied, and 2) all
energy level spacings change in the same ratio
in the thermally isolated process \cite{adiabatic}. Because this is the quantum counterpart
of classical adiabatic process (thermally isolated plus quasistatic process) \cite{adiabatic}.
Besides, the ratio of the change of the energy level spacings in the quantum adiabatic
process $D\rightarrow A$ should be equal to the ratio $\frac{T_{A}}{T_{D}}$ of the temperatures of the heat bath at $A$ and at $D$ (see Fig. 5); 
the energy level spacing at $C$ should be equal to that at point $D$ (see Fig. 5).

\begin{figure}[ht]
\begin{center}
\includegraphics[width=8cm, clip]{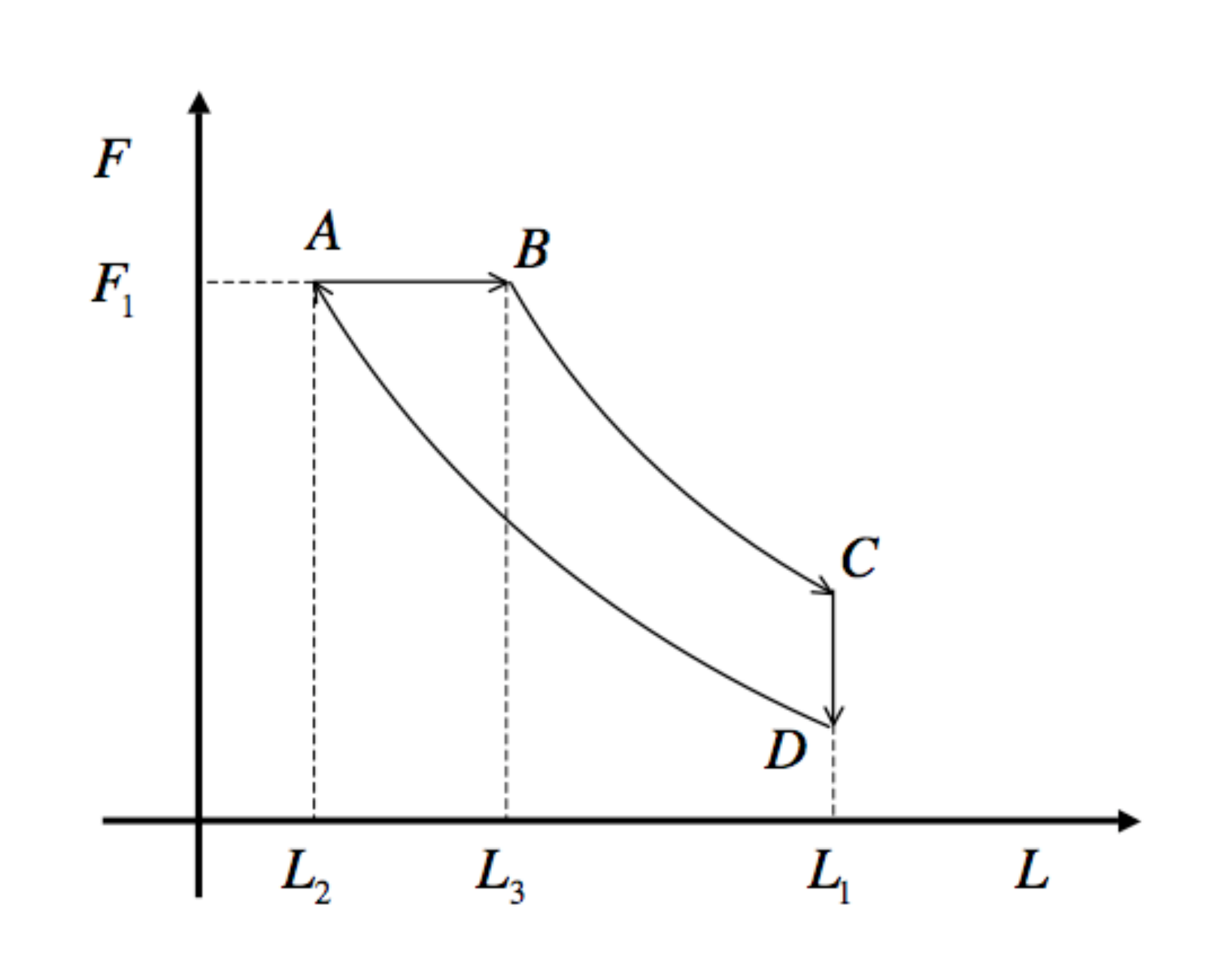}
\end{center}
\caption{Force-Displacement $F-L$ diagram of a quantum Diesel cycle
based on 1DB and single mode radiation field. $A\longrightarrow B$
represents an isobaric expansion process with a constant pressure
$F_1$; $B\longrightarrow C$ represents an adiabatic expansion
process with constant entropy; $C\longrightarrow D$ represents an
isochoric compression process with constant volume $L_1$;
$D\longrightarrow A$ is another adiabatic compression process.}
\label{fig6}
\end{figure}
In the following we will consider implementing the quantum Diesel cycle
in 1DB and in 1DH. First we consider the 1DB. 
The efficiency of a quantum Diesel cycle based on 1DB can be obtained through a straightforward
calculation (see Appendix B)
\begin{equation}
\eta_{\mathrm{Diesel}}=1-\frac{1}{3}\frac{r_{E}^{3}-r_{C}^{3}}{r_{E}-r_{C}}=1-\frac{1}{3}(r_{E}^{2}+r_{C}r_{E}+r_{C}^{2}).
\end{equation}
Here $r_{C}\equiv \frac{L_{2}}{L_{1}}$ (see Fig. 5) and $r_{E}\equiv \frac{L_{3}}{L_{1}}$ (see Fig. 5) are the compression 
and expansion ratios of the volumes. This efficiency for a quantum Diesel cycle based on 1DB agrees with
that of a classical Diesel cycle, too. Through a similar analysis we
obtain the efficiency for a quantum Diesel cycle based on 1DH with
the only change of $\gamma$ from 3 in Eq. (15) to 2
\begin{equation}
\eta^{\prime}_{\mathrm{Diesel}}=1-\frac{1}{2}\frac{r_{E}^{2}-r_{C}^{2}}{r_{E}-r_{C}}=1-\frac{1}{2}(r_{E}+r_{C}).
\end{equation}

\begin{table*}[tbp]
\caption{Working efficiencies of typical classical thermodynamic
cycles and their quantum counterparts based on different kinds of working substance. It can be seen that 1) except
the Carnot cycle, the efficiencies of all the thermodynamic cycles
are working substance-dependent, and 2) both quantum thermodynamic
cycles and classical thermodynamic cycles have the same efficiency as long as the adiabatic exponent is the same.
Adiabatic exponents for monoatomic, diatomic, and polyatomic
classical idea gas can be found in \cite{adiabatic exponent}. Here
$T_{C}$ and $T_{H}$ are the temperatures of the cold and hot
reservoirs; $V_{0}$ ($L_{0}$, $S_{0}$) and $V_{1}$ ($L_{1}$,
$S_{1}$) are the volume (length, area) of the working substance in
two isochoric processes; $P_{0}$ ($F_{0}$) and $P_{1}$ ($F_{1}$) are
the pressure (force) of the working substance in the two isobaric
processes.}
\begin{center}
\begin{tabular}{c|c|c|c|c|c}
\hline\hline
\multicolumn{2}{ }{}&\parbox{2.2cm} {\ \ \textbf{Carnot} \ \ \ \ \ \ \ \ \ \ \ \ \ \ \ \ \ (two isothermal + two adiabatic)} & \parbox{2.5cm} {\ \ \textbf{Otto} \ \ \ \ \ \ \ \ \ \ \ \ \ \ \ \ \ (two isochoric+ two adiabatic)} & \parbox{2.5cm} {\ \ \textbf{Brayton} \ \ \ \ \ \ \ \ \ \ \ \ \ \ \ \ \ (two isobaric + two adiabatic)} &\parbox{3.5cm} {\ \ \textbf{Diesel} \ \ \ \ \ \ \ \ \ \ \ \ \ \ \ \ \ (isochoric + isobaric + two isobaric)} \\

\multicolumn{2}{ }{}&$\eta=1-\frac {T_C} {T_H}$ & $\eta=1- \left ( \frac {V_0} {V_1} \right )^{\gamma-1}$  & $\eta=1- \left ( \frac {P_0} {P_1} \right )^{1-\frac{1}{\gamma}}$ &$\eta=1- \frac{1}{\gamma}\frac{\left(\frac{V_{2}}{V_{1}}\right)^{\gamma}-\left(\frac{V_{3}}{V_{1}}\right)^{\gamma}}{\left(\frac{V_{2}}{V_{1}}\right)-\left(\frac{V_{3}}{V_{1}}\right)}$ \\
 \hline\hline

                      & Monoatomic classical idea gas ($\gamma=\frac{5}{3}$) &$\eta=1-\frac{T_C}{T_H}$ &$\eta=1- \left ( \frac {V_0} {V_1} \right )^{\frac{2}{3}}$ &$\eta=1- \left ( \frac {P_0} {P_1} \right )^{\frac{2}{5}}$ & $\eta=1- \frac{3}{5}\frac{\left(\frac{V_{2}}{V_{1}}\right)^{\frac{5}{3}}-\left(\frac{V_{3}}{V_{1}}\right)^{\frac{5}{3}}}{\left(\frac{V_{2}}{V_{1}}\right)-\left(\frac{V_{3}}{V_{1}}\right)}$\\

\cline{2-6}  \textbf{Classical} & Diatomic classical idea gas ($\gamma=\frac{7}{5}$) & $\eta=1-\frac {T_C} {T_H}$  &$\eta=1- \left ( \frac {V_0} {V_1} \right )^{\frac{2}{5}}$ &$\eta=1- \left ( \frac {P_0} {P_1} \right )^{\frac{2}{7}}$ & $\eta=1- \frac{5}{7}\frac{\left(\frac{V_{2}}{V_{1}}\right)^{\frac{7}{5}}-\left(\frac{V_{3}}{V_{1}}\right)^{\frac{7}{5}}}{\left(\frac{V_{2}}{V_{1}}\right)-\left(\frac{V_{3}}{V_{1}}\right)}$\\  \cline{2-6}

                     &Polyatomic classical idea gas ($\gamma=\frac{4}{3}$) &$\eta=1-\frac{T_C}{T_H}$ &$\eta=1- \left ( \frac {V_0} {V_1} \right )^{\frac{1}{3}}$ &$\eta=1- \left ( \frac {P_0} {P_1} \right )^{\frac{1}{4}}$ & $\eta=1- \frac{3}{4}\frac{\left(\frac{V_{2}}{V_{1}}\right)^{\frac{4}{3}}-\left(\frac{V_{3}}{V_{1}}\right)^{\frac{4}{3}}}{\left(\frac{V_{2}}{V_{1}}\right)-\left(\frac{V_{3}}{V_{1}}\right)}$\\
\hline\hline

                     &Single particle in 1D box ($\gamma=3$) &$\eta=1-\frac{T_C}{T_H}$ & $\eta=1-\left (\frac{L_0}{L_1}\right)^2 $ &$\eta=1-\left(\frac{F_0}{F_1}\right)^\frac{2}{3}$&$\eta=1- \frac{1}{3}\frac{\left(\frac{L_{2}}{L_{1}}\right)^{3}-\left(\frac{L_{3}}{l_{1}}\right)^{3}}{\left(\frac{L_{2}}{L_{1}}\right)-\left(\frac{L_{3}}{L_{1}}\right)}$\\

\cline{2-6}  &Single particle in 2D box ($\gamma=2 $) & $\eta=1-\frac {T_C} {T_H}$  &$\eta=1-  \frac {S_0} {S_1} $ &$\eta=1- \left ( \frac {P_0} {P_1} \right )^{\frac{1}{2}}$ & $\eta=1- \frac{1}{2} \left(\frac{S_{2}}{S_{1}}- \frac{S_{3}}{S_{1}}\right) $\\ \cline{2-6}

\cline{2-6}  &Single particle in 3D box ($\gamma=\frac{5}{3}$) & $\eta=1-\frac {T_C} {T_H}$  &$\eta=1- \left ( \frac {V_0} {V_1} \right )^{\frac{2}{3}}$ &$\eta=1- \left ( \frac {P_0} {P_1} \right )^{\frac{2}{5}}$ & $\eta=1- \frac{3}{5}\frac{\left(\frac{V_{2}}{V_{1}}\right)^{\frac{5}{3}}-\left(\frac{V_{3}}{V_{1}}\right)^{\frac{5}{3}}}{\left(\frac{V_{2}}{V_{1}}\right)-\left(\frac{V_{3}}{V_{1}}\right)}$\\ \cline{2-6}

\cline{2-6}  &  1D Single mode photon field ($\gamma=2$) & $\eta=1-\frac {T_C} {T_H}$  & $\eta=1- \frac{L_0}{L_1} $ &$\eta=1-\left(\frac{F_0}{F_1}\right)^\frac{1}{2}$&$\eta=1- \frac{1}{2}\left(\frac{L_{2}}{L_{1}}-\frac{L_{3}}{L_{1}} \right)$\\ \cline{2-6}

\cline{2-6}  \textbf{Quantum} & 3D Black body radiation field ($\gamma=\frac{4}{3}$) &  $\eta=1-\frac {T_C} {T_H}$ &$\eta=1- \left ( \frac {V_0} {V_1} \right )^{\frac{1}{3}}$ &$\eta=1- \left ( \frac {P_0} {P_1} \right )^{\frac{1}{4}}$ & $\eta=1- \frac{3}{4}\frac{\left(\frac{V_{2}}{V_{1}}\right)^{\frac{4}{3}}-\left(\frac{V_{3}}{V_{1}}\right)^{\frac{4}{3}}}{\left(\frac{V_{2}}{V_{1}}\right)-\left(\frac{V_{3}}{V_{1}}\right)}$\\ \cline{2-6}

\cline{2-6}  & 1D harmonic oscillator ($\gamma= 2 $) & $\eta=1-\frac {T_C} {T_H}$  & $\eta=1- \frac{L_0}{L_1} $ &$\eta=1-\left(\frac{F_0}{F_1}\right)^\frac{1}{2}$&$\eta=1- \frac{1}{2}\left(\frac{L_{2}}{L_{1}}-\frac{L_{3}}{L_{1}} \right)$\\ \cline{2-6}

\cline{2-6} & 2D harmonic oscillator ($\gamma= \frac{3}{2}$) & $\eta=1-\frac {T_C} {T_H}$  &$\eta=1- \left ( \frac {S_0} {S_1} \right )^{\frac{1}{2}}$ &$\eta=1- \left ( \frac {P_0} {P_1} \right )^{\frac{1}{3}}$ & $\eta=1- \frac{2}{3}\frac{\left(\frac{S_{2}}{S_{1}}\right)^{\frac{3}{2}}-\left(\frac{S_{3}}{S_{1}}\right)^{\frac{3}{2}}}{\left(\frac{S_{2}}{S_{1}}\right)-\left(\frac{S_{3}}{S_{1}}\right)}$\\ \cline{2-6}

\cline{2-6} & 3D harmonic oscillator ($\gamma=\frac{4}{3}$) & $\eta=1-\frac {T_C} {T_H}$  &$\eta=1- \left ( \frac {V_0} {V_1} \right )^{\frac{1}{3}}$ &$\eta=1- \left ( \frac {P_0} {P_1} \right )^{\frac{1}{4}}$ & $\eta=1- \frac{3}{4}\frac{\left(\frac{V_{2}}{V_{1}}\right)^{\frac{4}{3}}-\left(\frac{V_{3}}{V_{1}}\right)^{\frac{4}{3}}}{\left(\frac{V_{2}}{V_{1}}\right)-\left(\frac{V_{3}}{V_{1}}\right)}$\\ \cline{2-6}

                   & spin-1/2 (2-level system) ($\gamma=2$) & $\eta=1-\frac {T_C} {T_H}$  & $\eta=1- \frac{L_0}{L_1} $ &$\eta=1-\left(\frac{F_0}{F_1}\right)^\frac{1}{2}$ & $\eta=1- \frac{1}{2}\left(\frac{L_{2}}{L_{1}}-\frac{L_{3}}{L_{1}} \right)$\\

 \hline\hline
\end{tabular}
\end{center}
\end{table*}

Before concluding this section, we would like to mention that we can
also discuss the quantum Brayton cycle and the quantum Diesel cycle
based on an arbitrary quantum system, such as the 3D black body
radiation field or a spin-1/2 system in an external magnetic field with the
Hamiltonian $H=\frac{1}{2}B \sigma_{z}$. Here $\sigma_{z}$ is the
Pauli matrix and $B$ is the external magnetic field. It can be seen from the Table II
that the efficiencies for both quantum Carnot cycle and classical Carnot cycle 
are always equal to the Carnot efficiency $1-\frac{T_{C}}{T_{H}}$ irrespective of the properties of the 
working substance. Different from
the Carnot cycle, the efficiencies of Otto Cycle, Brayton cycle and
Diesel cycle are working substance-dependent (see Table II). More
specifically, they depend on the adiabatic exponent $\gamma$ of the
working substance. As long as we get the adiabatic exponent
$\gamma$ of the quantum system, we obtain the explicit expression of the efficiencies of
the quantum thermodynamic cycles by substituting $\gamma$ into the expression
of the efficiencies of the classical thermodynamics with their
adiabatic exponent. For example, for a spin-1/2 system in an external magnetic field, we choose the inverse of the
magnetic field strength as the generalized coordinate
$L=\frac{1}{B}$. Then it can be found that the adiabatic exponent
for such a system is $\gamma=2$. As a result, the efficiencies of a
quantum Brayton cycle and a quantum Diesel cycle based on a spin-1/2 system in an
external magnetic field are the same as that based on 1DH (14), (16). Similarly, the
efficiencies for a Brayton cycle and a Diesel cycle based on 3D radiation
field can be obtained straightforwardly by substituting $\gamma$
with the adiabatic exponent $\frac{4}{3}$ \cite{fourprocesses,atoz} for 3D radiation field.
In Table II we list the working efficiencies for several typical
thermodynamic cycles based on different kinds of classical and
quantum working substances (based on their adiabatic exponent).

\section{CONCLUSIONS AND REMARKS}

In summary, in this paper, we study the quantum mechanical analogy
of the classical isobaric process based on a microscopic definition
of force. In studying the thermodynamic properties of a small
quantum system, we use a new pair of conjugate variables $P_{n}$ and $E_{n}$ instead of the usual thermodynamic
variables $P$ and $V$ or $T$ and $S$ \cite{conjugate}. The general expression of force for an
arbitrary quantum system $F=-\sum_n P_{n}\frac{dE_n(L)}{dL}$ is
found. It can be checked that this expression is in accordance with
the force $F=-\left(\frac{\partial \mathbb{F} }{\partial
L}\right)_{T} $ in statistical mechanics if the quantum system is in thermal equilibrium with a heat bath.
In addition we clarify the relation between adiabatic process (thermally isolated plus quasistatic process)
in classical systems and quantum adiabatic process in quantum systems, 
and we find that all energy level spacings change in the same ratio in the quantum adiabatic
process is essential in simulating the classical adiabatic process. Otherwise irreversibility will
arise \cite{adiabatic}. Based on quantum isobaric processes, we make quantum mechanical extension of some typical
thermodynamic cycles. The properties of these quantum thermodynamic processes and cycles are
clarified, and we bridges the quantum thermodynamic cycles and their classical counterpart.
The quantum heat engines and their classical counterparts have the same
efficiencies as long as their working substance has the same
adiabatic exponent. The definitions of force and work
for a single-particle quantum system may have important application
in the experimental exploration of nonequilibrium thermodynamics in
small quantum systems, such as quantum Jarzynski equality and
quantum Crooks Fluctuation Theorem \cite{phystoday05,quan08}. 
Though the working substance of quantum heat
engines deviates from thermodynamic limit, we reproduce the
efficiency of classical heat engines. Hence our study  lay the
concrete foundation for Szilard-Zurek single molecule engine. Moreover, we found the close relation between classical ideal gas
and 1DB, and between single-mode photon gas and 1DH.

Before concluding this paper, we would like to mention that in our
current study we focus on the quantum single-particle system, and
its related quantum mechanical generalization of heat, work,
pressure, and we regain the results of classical thermodynamic
processes and cycles. We also notice some studies about quantum heat
engines with quantum many body system as the working substance
\cite{manybody}. For quantum many body system, e.g., ideal bosonic
gas or ideal fermionic gas, the mechanical variable, such as heat
work, pressure, are well defined and their equation of state as well
as their expression of internal energy \cite {pathria} deviate from
that of the classical ideal gas. As a result, the properties of
quantum thermodynamic cycles based on the quantum many-body system
deviate from that of classical ideal gas due to quantum degeneracy.
Finally, similar to the discussion about finite-power Carnot engine
\cite{curzon}, we can discuss about finite-time quantum Brayton
cycle and quantum Diesel cycle. Finite-power analysis of Brayton
cycle and Diesel cycle will be given later.

\section{acknowledgments}
This work is supported by U.S. Department of Energy through the
LANL/LDRD Program. The author thanks Prof. P. W. Milonni for helpful discussions and Prof. J. Q. You for the hospitality
extended to him during his visit to Fudan university.

\begin{appendix}

\setcounter{section}{0} \setcounter{equation}{0} \renewcommand{\thesection}{%
\Alph{section}}

\appendix
\section{OPERATION EFFICIENCY OF QUANTUM BRAYTON CYCLE}

According to the
definition of heat exchange (\ref{1}) in the quantum mechanical
system, we obtain the heat absorbed by the system from a
time-dependent heat bath during the quantum isobaric expansion
process $A\longrightarrow B$ (see Fig. 3 and Fig. 4)
\begin{equation}
\begin{split}
\dbarit Q_{AB}=&\int_{L_A}^{L_B}\left[\sum_nE_n(L)\frac{dP_n(L)}{dL}\right]dL\\
=&\sum_n\int_{L_A}^{L_B}\left[ \left[E_n(L)P_n(L)\right]^{\prime}-\frac{dE_n(L)}{dL}P_n(L)\right] dL \label {14}\\
=&\sum_n[E_n(L_B)P_n(L_B)-E_n(L_A)P_n(L_A)]\\
&+\int_{L_A}^{L_B}F(L)dL\\
=&\frac{1}{2}[F_1L_B-F_1L_A]+F_1(L_B-L_A)\\
=&\frac{3}{2}F_1(L_B-L_A).
\end{split}
\end{equation}
In obtaining the above result we have used Eq. (5) and Eq. (10).
Similarly, we obtain the heat released to the time-dependent entropy
sink
\begin{equation}
\dbarit Q_{CD}=\frac{3}{2}F_0(L_C-L_D). \label {15}
\end{equation}
Hence, the efficiency of the quantum Brayton cycle based on a 1DB
can be expressed as
\begin{equation}
\eta_{\mathrm{Brayton}}=1-\frac{F_0(L_C-L_D)}{F_1(L_B-L_A)}. \label {16}
\end{equation}
Due to the equation of motion (5) and the expression of the internal
energy (10), we have $F_1 \times L_B/2=U(L_B)$, $F_0 \times
L_C/2=U(L_C)$. In addition to the relation of the internal energies
in the quantum adiabatic process $B\longrightarrow C$
\begin{equation}
\frac{U(L_B)}{U(L_C)}=\left(\frac{L_C}{L_B}\right)^2, \label {17}
\end{equation}
we have
\begin{equation}
\frac{F_1}{F_0}=\left(\frac{L_C}{L_B}\right)^3 \label {18}
\end{equation}
for the quantum adiabatic process $B\longrightarrow C$.
Through a similar analysis we obtain
\begin{equation}
\frac{F_1}{F_0}=\left(\frac{L_D}{L_A}\right)^3 \label {19}
\end{equation}
for another quantum adiabatic process $D \longrightarrow A$. Based
on all the above results (A3), (A5), and (A6), we obtain the
efficiency of the quantum Brayton cycle based on the 1DB
\begin{equation}
\eta_{\mathrm{Brayton}}=1-\left(\frac{F_0}{F_1}\right)^{\frac{2}{3}}. \label {20}
\end{equation}

In the following we consider a quantum Brayton cycle based on 1DH. Similar to the above analysis, we calculate the heat absorbed by the
system during the quantum isobaric expansion process $A
\longrightarrow B$ (see Fig. 3)
\begin{equation}
\begin{split}
\dbarit Q_{AB}=&\int_{L_A}^{L_B}\left[\sum_nE_n(L)\frac{dP_n(L)}{dL}\right]dL \label {24}\\
=&[U(L_B)-U(L_A)]+\int_{L_A}^{L_B}F d(L)\\
=&\left(\frac{\hbar\omega_B}{e^{\beta(L_B) \hbar\omega_B}-1}+\frac{\hbar\omega_B}{2}\right)\\
&-\left(\frac{\hbar\omega_A}{e^{\beta(L_A) \hbar\omega_A}-1}+\frac{\hbar\omega_A}{2}\right)+F_H (L_B-L_A)\\
=&F_1 (L_B-L_A),
\end{split}
\end{equation}
where we have used the relations (7) and (12) in the quantum
isobaric process ($A \longrightarrow B$). Similarly, we obtain the
heat released to the entropy sink in another quantum isobaric
process $C \longrightarrow D$
\begin{equation}
\dbarit Q_{CD}=F_0 \left(L_{C}-L_{D}\right). \label {27}
\end{equation}
The efficiency of the quantum Brayton cycle based on a 1DH can be
expressed as
\begin{equation}
\eta^{\prime}_{\mathrm{Brayton}}=1-\frac{F_0(L_{C}-L_{D})}{F_1(L_{B}-L_{A})}.\label{eta}
\end{equation}
From Eqs. (7) and (12) we have $F_1\times L_{B}=U(L_{B})$ and
$F_0\times L_{C}=U(L_{C})$. In addition to the relation of internal
energy in the quantum adiabatic process
\begin{equation}
\frac{U(L_{B})}{U(L_{C})}=\frac{L_{C}}{L_{B}}, \label {29}
\end{equation}
we have
\begin{equation}
\frac{F_1}{F_0}=\left(\frac{L_{C}}{L_{B}}\right)^2. \label {30}
\end{equation}
Hence, from Eqs. (A10) and (A12) we obtain the efficiency of a quantum
Brayton cycle based on 1DH
\begin{equation}
\eta^{\prime}_{\mathrm{Brayton}}=1-\sqrt{\frac{F_0}{F_1}}. \label {31}
\end{equation}

\section{OPERATION EFFICIENCY OF QUANTUM DISEL CYCLE}

For a quantum Diesel cycle (see Fig. 5), the input energy in
the quantum isobaric process $A\rightarrow B$ and the output energy in the quantum
isochoric process $C\rightarrow D$ can be calculated as
\begin{equation}
\begin{split}
Q_{in}&=C_{P}(T_{B}-T_{A}),\\
Q_{out}&=C_{V}(T_{C}-T_{D}),
\end{split}
\end{equation}
where $C_{P}$ and $C_{V}$ are the heat capacity at constant pressure
and constant volume respectively; $T_{A}$, $T_{B}$, $T_{C}$, and
$T_{D}$ are the temperatures of the system at different points of the Diesel cycle (see
Fig. 5). Thus the efficiency of the quantum Diesel cycle can be
expressed in terms of heat capacities and temperatures
\begin{equation}
\eta=\frac{Q_{in}-Q_{out}}{Q_{in}}=1-\frac{C_{V}(T_{C}-T_{D})}{C_{P}(T_{B}-T_{A})}.
\end{equation}
It is convenient to express this efficiency (B2) in terms of
compression ration $r_{C}\equiv \frac{L_{2}}{L_{1}}$ (see Fig. 5)
and the expansion ratio $r_{E}\equiv \frac{L_{3}}{L_{1}}$ (see Fig.
5) of the volumes. Now using the equation of state $FL=kT$ (5) and
$\frac{C_{P}}{C_{V}}= \gamma=3$ for 1DB, the efficiency (B2) can be
rewritten as
\begin{equation}
\eta=1-\frac{1}{3}\frac{(F_{C}L_{C}-F_{D}L_{D})}{(F_{B}L_{B}-F_{A}L_{A})}.
\end{equation}
By utilizing the facts $L_{C}=L_{D}=L_1$ and $F_{A}=F_{B}=F_{1}$ (see Fig. 5), we further simplify the Eq. (B3) to
\begin{equation}
\eta=1-\frac{1}{3}\frac{L_{1}(F_{C}-F_{D})}{F_{1}(L_{B}-L_{A})}=1-\frac{1}{3}\frac{(\frac{F_{C}}{F_{1}}-\frac{F_{D}}{F_{1}})}{(r_{E}-r_{C})}.
\end{equation}
Finally by making use of the adiabatic condition $F L^{3}=const$ for
1DB in the quantum adiabatic process, we obtain
\begin{equation}
\begin{split}
\frac{F_{C}}{F_{1}}&=\left(\frac{L_{3}}{L_{1}}\right)^{3}=r_{E}^{3},\\
\frac{F_{D}}{F_{1}}&=\left(\frac{L_{2}}{L_{1}}\right)^{3}=r_{C}^{3}.
\end{split}
\end{equation}
Substituting Eq. (B5) into Eq. (B4), the efficiency of a quantum Diesel cycle based on 1DB can be written
as
\begin{equation}
\eta_{\mathrm{Diesel}}=1-\frac{1}{3}\frac{r_{E}^{3}-r_{C}^{3}}{r_{E}-r_{C}}=1-\frac{1}{3}(r_{E}^{2}+r_{C}r_{E}+r_{C}^{2}).
\end{equation}

\end{appendix}


\begin{thebibliography}{99}
\bibitem{casimir} Here we consider the zero point energy of the specific mode $\omega$ of the radiation field inside the cavity (6). To be more accurate we should also consider the zero point energies of the remaining permissible discrete modes inside the cavity
and the continuous modes outside the cavity as well. The net effect of all these zero point energies is the Casimir force \cite{milonni}, and it is a tiny correction to the usual radiation pressure. When taking into account of all those modes, the value of the correction
should be $-\frac{\hbar \pi c}{12 L^{2}}$  \cite{milonni} instead of $\frac{s \hbar \pi c}{2 L^{2}}$ in Eq. (7). Here the negative and positive signs mean attractive and repulsive force respectively. Nevertheless the Casimir effect is not our focus and it does not affect our main result. So we do not include this effect in our discussion.
\bibitem{adiabatic} In the classical thermodynamic cycles, the adiabatic process implies 1) thermally isolated and 2) quaistatic, or very slow \cite{fourprocesses,generalizedforce}.
If the thermally isolated process proceeds fast, the quasistatic conditions are not satisfied. As a result, there will be internal entropy increase, and irreversibility arises. In order to construct a reversible cycle, such as 
Carnot cycle and Brayton cycle, quasistatic condition must be satisfied in the thermally isolated process. Similarly. for a quantum thermodynamic cycle, the quantum mechanical counterpart of classical
adiabatic process implies 1) thermally isolated and 2) no interstate excitations \cite{quan07} and 3) all energy level spacings change in the same ratio. Alternatively, the quantum mechanical counterpart of
adiabatic process is quantum adiabatic process given that all energy level spacings change in the same ratio. This is because if one starts from an Gibbs density operator, when the quantum adiabatic conditions are satisfied 
and all energy level spacings change in the same ratio as we change a parameter of the Hamiltonian,
it will remain Gibbsian. So there is no source of irreversibility. On the contrary, if either 1) the quantum adiabatic conditions are not satisfied, or 2) the energy level spacings do not change in the same ratio in the thermally isolated process,
the density operator will not remain Gibbsian, and irreversibility arises. Hence, in order to construct a reversible quantum thermodynamic cycles (quantum mechanical counterpart of reversible
classical thermodynamic cycles), we must focus on those systems whose energy level spacings change in the same ratio and use quantum adiabatic process to replace the classical adiabatic plus quasistatic process.
\bibitem{conjugate} There are many other types of conjugate pair of variables, for example, the electromotive force and the amount of charge, the magnetization and the magnetic field, the surface tension and the surface area, the elastic force and the length stretched, and the gravitational potential and the mass.
\end{thebibliography}
\end{document}